\documentstyle[amssymb,aps,twocolumn]{revtex}
\begin{document}
\author{J. Bobroff$^{1}$, H. Alloul$^{1}$, S.\ Ouazi$^{1}$, P.\ Mendels$^{1}$, A.\
Mahajan$^{1\ast }$, N.\ Blanchard$^{1}$, G.\ Collin$^{2}$, V.\ Guillen$^{3}$%
, J.-F.\ Marucco$^{3}$}
\title{Absence of static phase separation in the high T$_{c}$ cuprate YBa$_{2}$Cu$%
_{3}$O$_{6+y}$}
\address{$^{1}$Laboratoire de Physique des Solides, UMR 8502, Universit\'{e}\\
Paris-Sud, 91405 Orsay, France\\
$^{2}$LLB, CE-Saclay, CEA-CNRS, 91191 Gif sur Yvette, France\\
$^{3}$LEMHE, UMR 8647, Universit\'{e} Paris-Sud, 91405 Orsay, France}
\date{\today}

\twocolumn[\hsize\textwidth\columnwidth\hsize\csname@twocolumnfalse\endcsname
\maketitle

\begin{abstract}
We use $^{89}$Y\ NMR in YBa$_{2}$Cu$_{3}$O$_{6+y}$ in order 
to evaluate with high sensitivity the distribution of hole content $p$ in the CuO$_{2}$ planes.
 For $y=1$ and $y=0.6$, this hole doping 
distribution is found narrow with a full width at half maximum smaller than $\Delta p=0.025$.
 This rules out any 
large static phase separation between underdoped and optimally doped
 regions in contrast 
with the one observed by STM in Bi2212 and by NQR in LaSrCuO. 
This establishes that static electronic phase separation is not a generic feature of the cuprates.
\end{abstract}

]

In the cuprates, the hole doping of the CuO$_{2}$ planes induces the unusual
features of the phase diagram: high-$T_{c}\;$superconductivity, strange
metal behavior, or pseudogap. A large body of theoretical work argues that
this hole doping could be intrinsically strongly inhomogeneous in the
planes, forming segregated hole-rich and hole-poor regions on a nanoscale 
\cite{emery}. This phase separation has been proposed to be essential to
explain the unusual properties of the cuprates, appearing for example as
stripes as argued from neutron scattering experiments \cite{NeutronsStripes}%
.\ Such proposals have been recently highlighted by STM\ studies which
reveal strong inhomogeneities of the superconducting properties at the
surface of Bi2212 \cite{ChangPRB92}\cite{Cren}\cite{Pan}\cite{howald}\cite
{LangSTM}. The Berkeley STM group have imaged a spatial distribution of the
superconducting gap which they associated with a distribution of
concentration of holes ranging from $p=0.1$ to $p=0.2$ holes/planar unit
cell \cite{Pan}.

The archetype of the cuprate families is YBa$_{2}$Cu$_{3}$O$_{6+y}$
(YBaCuO). If these electronic inhomogeneities do not exist in this YBaCuO
family, then they result of specific disorder in other compounds. However,
STM measurements are not as extensive in YBaCuO due to surface cleaving
problems and oxygen loss in vacuum. The existing STM studies display an
inhomogeneous or relatively homogeneous surface depending on the surface
preparation procedure \cite{edwards}\cite{yeh}.\ These limitations do not
occur if one uses other local probes such as nuclear magnetic (NMR) or
quadrupolar resonance (NQR).\ The huge body of NMR/NQR studies done so far
concentrated on the $p-$variation of the average values of specific
quantities (i.e. the static NMR\ shift $K$\ or the relaxation time $T_{1}$%
).\ However, as will be emphasized hereafter, the NMR/NQR spectroscopy also
allows to determine the distribution of these quantities in the{\em \ bulk }%
samples.\ NMR is then sensitive to the local distribution of electronic
properties like STM, but not to the specificities of the surface. As an
appealing example, Singer {\it et al } \cite{Singer} evidenced recently a
distribution of $T_{1}$ over the Cu NQR spectrum in bulk LaSrCuO, which can
be attributed to a distribution of $p$ as large as the one observed on the
Bi2212\ surface.\ 

We present an NMR static study of the YBaCuO family using $^{89}$Y\ NMR
spectra.\ This allows us to estimate the shape of the hole distribution in
the bulk{\em \ }of the compound. We study two doping compositions
(underdoped O$_{6.6}$ and slightly overdoped O$_{7}$).\ At O$_{7}$ the
oxygen chains reservoirs are full, thus well ordered.\ At O$_{6.6}$, the
various properties are weakly oxygen content dependent.\ So both
compositions should produce a minimal disorder of doping in the planes. We
show indeed that our results reveal a very narrow hole distribution. These
results will be compared with other experiments in the various cuprate
families.

We synthesized a large sample batch of single crystal grains of YBaCuO.\ It
was oxidized at $T=350%
%TCIMACRO{\UNICODE[m]{0xb0}}%
%BeginExpansion
{{}^\circ}%
%EndExpansion
C$, and then slowly cooled under oxygen atmosphere from 350$%
%TCIMACRO{\UNICODE[m]{0xb0}}%
%BeginExpansion
{{}^\circ}%
%EndExpansion
C$ to 270$%
%TCIMACRO{\UNICODE[m]{0xb0}}%
%BeginExpansion
{{}^\circ}%
%EndExpansion
C$ $\ $at a rate of $0.4K$/hr.\ After this procedure, the sample has a
nominal maximum oxygen content $y=1\;$with a reduced $T_{c}=89.1$ K with
respect to optimal doping ($T_{c}=92.5$K). A part of this batch was reduced
to $y=0.60\pm 0.02$ ($T_{c}=55.8$K) under primary vacuum ($P=0.13$ mbar) at $%
T=385%
%TCIMACRO{\UNICODE[m]{0xb0}}%
%BeginExpansion
{{}^\circ}%
%EndExpansion
C$ . The oxygen content was measured by thermogravimetry, i.e.\ measurement
of weight loss during the deoxidisation process until equilibrium is reached
in the above conditions. For both samples, the crystallites were then
aligned in stycast epoxy in the field $H_{ext}(\simeq 7.5$ Tesla) of the NMR
spectrometer \cite{FarellPRB}. This field corresponds to a reference NMR
resonance $^{89}\nu =15634.67$ kHz for a liquid YCl$_{3}$ solution. Spectra
were obtained for H$_{ext}$ perpendicular to the $c$ cristallographic axis
using a standard $\pi /2-\pi $ pulse sequence and Fourier transform of half
of the spin echo \cite{footnoteAlignement}.\ {\em \ }The repetition time of
the pulse sequence was long enough at each temperature to recover the
saturation signal for any value of $p$.\ For a given T, the maximum of T$%
_{1}(p)$ varies from $17$ to $110$ sec when decreasing $T$ from $300K$ down
to $120K$ \cite{Alloul89}.

$^{89}$Y\ NMR probes\ sensitively any doping distribution in the CuO$_{2}$
planes because the Y nucleus is coupled to Cu of its adjacent CuO$_{2}$
planes through the oxygen orbitals \cite{Alloul89}.\ This results in a shift
of the NMR line related to the Cu spin susceptibility $\chi $ given by

\begin{equation}
^{89}K(T)=\frac{8A_{hf}\chi (T)}{\mu _{B}}+^{89}\delta  \label{formulaShift}
\end{equation}

{\em Here, }$^{89}K(T)${\em \ is strongly dependent on hole content because
of the strong hole doping dependence of }$\chi (T)$. In contrast, the $T-$%
independent chemical shift $^{89}\delta $ \ and hyperfine coupling $%
A_{hf}=-1.95$ $kOe/\mu _{B}$ between the $^{89}Y$ and one Cu do not change
much with hole doping ($\mu _{B}\;$is the Bohr magneton) \cite{Alloul89}.
This relation is illustrated in Fig.1 which displays the shift of the peak
value of the $^{{\it 89}}$Y\ NMR\ line. It has been taken for a series of
samples prepared from a single batch for different $y$. The shift $^{89}K(T)$
displays the well known behaviour for $\chi (T)$.\ It is nearly constant
above optimal doping.\ It displays a marked decrease at low T which is the
signature of the pseudogap for the underdoped cases. The oxygen content $y$
can be converted into the hole content per plane using the parabolic
dependence of $T_{c}$ \cite{TallonPRB95} given by $p=0.16\pm \sqrt{%
(1-T_{c}(y)/T_{c}^{\max })/82.6}$. The corresponding $p$ and experimental $%
T_{c}$ are plotted in the inset of fig.1. Note that near optimal doping, the
large $p$ dependence of $^{89}K$ makes it a sensitive probe to any doping
variation: for example, at $T=120$\ K, a change in $p$ as small as $0.01$
corresponds to a change in $^{89}K$ of about 20 ppm, easily detectable. In
the underdoped regime, the sensitivity to the doping is smaller but still
sizeable.

The NMR spectrum of a given sample consists of a histogram of the NMR\
shifts throughout the sample.\ Note that, as STM, the NMR local probe is
sensitive to variations nearly on the atomic scale, as each Y nucleus is
coupled only to its 8\ nearest neighbour Cu sites. Then, from the
correspondence shift-doping of fig.1, any local distribution of doping will
lead to a similar distribution in the shift, hence to a broadening of the
spectrum.

Let us now present the experimental $^{89}$Y\ NMR spectra for $y=0.6$ and $1$
from which we will extract quantitatively the actual doping distribution.
The spectra are plotted for $T=120$ K and $300$ K for the optimally and
underdoped samples in fig.2. As can be observed immediately there is no
significant overlap between the two spectra at $T=120$ K . This already
proves that the O$_{7}$ ($p=0.18$) sample contains no appreciable amount of
underdoped nanoscale regions (equivalent to $p=0.1$) and vice versa. In
fig.\ 2 we also display the hole doping scales deduced from the relation
between $p$ and $^{89}K$ obtained from fig.1.{\em \ }From these hole doping
scales it can be seen roughly that the full width at half maximum FWHM of
the doping distribution is necessarily smaller than $\Delta p\simeq 0.04$ at
optimal doping and $0.02$ for the underdoped case.\ This clearly proves
without any further analysis that the observed doping distribution in YBaCuO
is much smaller than in Bi2212 and LaSrCu3O where $\Delta p\simeq 0.1$.

For a more quantitative analysis, we need to consider besides the
distribution of $\chi (T)$ any distribution of the chemical shift $\delta $
and hyperfine coupling $A_{hf}\;$which enter Eq.\ref{formulaShift}. Such
distributions will come from any local structural disorder. These
distributions are fully responsible for the FWHM $\Delta K=32\pm 5$ $ppm$ of
the undoped compound for $T>T_{N}$ \cite{Alloul1990}. Indeed, for oxygen
content $y<0.15$, no hole distribution broadening is expected \cite
{footnote1}.\ This results from the fact that xoxygens introduced in the Cu
reservoir layer mainly convert the two adjacent 3d$^{10}$\ Cu(1) into 3d$^{9}
$ and do not yield any hole doping of the CuO$_{2}$ planes, so that $p$ is
strictly $0$. It is confirmed by the fact that the spectrum has the
narrowest width amoung all dopings and is not found significantly dependent
on oxygen content up to $y=0.15$.

For higher $y$, if $\Delta \delta $ and $\Delta A_{hf}$ are the FWHM of the
respective distributions assumed to be uncorrelated and gaussian for
simplicity, this will lead to an additional broadening with FWHM $\Delta
K(y,T)$:

\begin{equation}
\Delta K^{2}(y,T)=\Delta \delta ^{2}(y)+\Delta K_{hf}^{2}(y,T)
\end{equation}
where $\Delta K_{hf}=8\mu _{B}^{-1}\Delta A_{hf}\chi (T)$. We note that the $%
y=0$ compound is the more ordered as it is tetragonal with no twin
boundaries and empty chains, hence $\Delta \delta $ and $\Delta A_{hf}$
should increase at higher $y$. As these broadenings add to the doping
induced broadening, taking their $y=0$ estimates at higher dopings will lead
to an {\em overestimate} of the doping distribution. As $\Delta K_{hf}$ is
T-dependent, we can evaluate $\Delta A_{hf}$ and $\Delta \delta $
separately. Indeed, at $y=0.6$, $\chi (T)$ doubles between 120 K\ and 300 K
so that $\Delta K_{hf}$ doubles as well, whereas the FWHM increases only by $%
11\%$ as seen in fig.2. As this T-dependence might as well be due to the
doping distribution, this leads to an upper bound $\Delta
A_{hf}/A_{hf}\leqslant 0.13$. Such an upper bound would be explained in a
naive hybridization computation by a random displacement of Cu by 0.04 \AA\
from its ideal position, a typical value from Rietveld measurements. We will
consider the two extreme cases:

({\it i}) $\Delta A_{hf}/A_{hf}=0$ corresponding to $\Delta \delta =32$ ppm

({\it ii}) $\Delta A_{hf}/A_{hf}=0.13$ corresponding to $\Delta \delta =29$
ppm.\ 

In order to extract the actual doping distribution, we start from a
distribution of oxygen doping $P(y)$. We convert it for different
temperatures into a shift distribution through the phenomenological $^{89}K$
vs $y$ variation deduced from Fig.1. We convolute this $K$ distribution with
the $\delta $ and $A_{hf}$ gaussian distributions with FWHM either {\it (i) }%
or{\it \ (ii)}. This process is iterated until an optimal $P(y)$ is found to
fit the experimental spectra {\em for all temperatures} with no additional
free parameter.

In the underdoped $y=0.6$ compound, this fitting procedure is limited by the
fact that the NMR shift is nearly insensitive to the doping distribution for 
$y<0.6$ as seen in fig.1.\ However, the thermogravimetry procedure during
deoxydation constrains the measure of the average oxygen level $\overline{y}%
=0.60\pm 0.02$. We thus assume $P(y)$ to be symetric around $\overline{y}$.\
The best fit is then obtained for a gaussian distribution $P(y)=\exp \left(
-(y-0.6)^{2}/\sigma ^{2}\right) $ with $\sigma =0.05$ for {\it (i)} and $%
\sigma =0.1$ for {\it (ii)}. This oxygen distribution $P(y)$ and the
corresponding hole distribution ${\cal P}(p)$ are plotted in fig.3 together
with the usual $T_{c}$ diagram. The corresponding simulations fit perfectly
the experimental ones at all temperatures (examples of fits are given as
dotted lines in fig.2).

For YBaCuO$_{7}$, the chains are completely full.\ Further oxidation is
prohibited. Therefore the distribution $P(y)$ is expected to be non
symmetric \cite{FootnoteOverdoping}.\ To take this limitation into account,
we model $P(y)$ as a convolution of a gaussian by $\exp (-\left| y-1\right|
/\lambda _{\pm })$ where $\lambda _{\pm }$ are allowed to differ for $y>1$
and $y<1$. The best fit is found for $\sigma =0.01,$ $\lambda _{-}=0.06,$
and $\lambda _{+}=0.025$ {\it (i) }or $0.004$ {\it (ii)}.\ $P(y)$ and the
corresponding fit of the spectra are respectively plotted in fig.3\ and
fig.2.\ The high asymmetric shape towards $y<1$ found for $P(y)$ confirms
that no large overdoping of the planes is produced locally, as expected. In
order to illustrate the high accuracy of our method, we choose $P(y)$
slightly broader than our best fit, plotted as the dotted line in the upper
panel of fig.3. The corresponding simulation plotted as a dashed line in the
upper panel of fig.2 clearly fails to fit the experimental spectrum.

In summary, our results demonstrate that the maximum possible distribution
of doping is quite sharp with typical width $\Delta p\leqslant 0.025$ for
optimal doping and $\Delta p\leqslant 0.01$ for the underdoped sample. The
distribution is much smaller than in the LaSrCuO and Bi2212 cuprates. In
LaSrCuO, at optimal doping, Singer {\it et al.} find $\Delta p\simeq 0.09$
as figured by the arrow in fig.3 \cite{Singer}. Another NMR\ study \cite
{Haase} using $^{17}$O\ and $^{63}$Cu NMR allowed to evidence a short length
scale spatial modulation in underdoped LaSrCuO compatible with Ref.\cite
{Singer}. In Bi2212, the STM experiments reveal regions on the surface with
the usual superconducting gap while others display a pseudogap \cite
{ChangPRB92}\cite{Cren}\cite{Pan}\cite{howald}\cite{LangSTM}. This has been
interpreted to be due either to some disorder effect \cite{Cren} or to
superconducting and insulating-like underdoped regions \cite{Pan}\cite
{howald}\cite{LangSTM}. The corresponding doping distribution of width $%
\Delta p=0.085$ extracted in Ref.\cite{Pan} is plotted in fig.3. Similar
widths are found for an optimally and an underdoped compound in Ref.\cite
{LangSTM}. Therefore, YBaCuO appears much more homogeneous than both LaSrCuO
and Bi2212. In other cuprates, no key experiment was performed to probe
locally the doping distribution, to our knowledge. However, in Hg1201 and
Tl2201, typical $^{17}$O NMR widths are similar to those in YBaCuO and much
smaller than in the Bi and La compounds \cite{bobroffPRL97}. Following the
analysis done above and the proportionality between $^{17}$O and $^{89}$Y
shifts, we then expect the Hg and Tl family not to exhibit any strong doping
distribution either.

In the LaSrCuO family, the large distribution seen by NQR occurs in the bulk
of the material and might be associated with tilt of the oxygen octaedra,
buckling of the planes, or stripes. In the Bi material, the STM results
might be only a surface specificity. In YBaCuO, the doping distribution $%
P(y) $ measured here is not only much narrower but might even have only a
macroscopic origin. As we used powders of cristallites with sizes smaller
than 30 $\mu m$, any oxygen gradient within each cristallite, or a
correlation between $\overline{y}$ and the cristallite size could lead to
the observed $P(y)$. At O$_{7}$, such oxygen gradients are explained by the
fact that oxygen diffusion between chains becomes limited at low
temperature. This naturally leads to the asymmetry of $P(y)$ towards $y<1$
seen in fig.3. This effect has been observed systematically in the various
studies published by our group using $^{89}$Y, or $^{17}$O NMR spectra in
many different YBCO$_{7}$ powders.\ At O$_{6.6}$, the observed broad $P(y)$
is very narrow when plotted versus $p$ (fig.3). Indeed a change in oxygen
content does not strongly modify the actual hole doping of the planes near $%
y=0.6$ hence leading to the plateau of $T_{c}$. This is to be associated
with the fact that extra oxygens at such composition occupy empty chains,
and do not modify the hole content, in analogy with the situation at $y=0$.
Therefore, the actual distributions would be much narrower on a submicron
size region similar to the one sampled by STM. For both dopings, the key
factor to the good homogeneity encountered is then probably the presence of
the chains. The existing chain disorder does not lead to a sizeable
distribution of hole content in the planes.\ Furthermore, those chains are
probably sufficiently far from the planes not to stabilize a charge
segregation. This specificity makes YBa$_{2}$Cu$_{3}$O$_{6.6\text{ or }7}$
some of the best prototypes of clean homogeneous cuprates.

In conclusion, the nanoscale static segregations observed so far cannot be
considered as an intrinsic phenomenon common to all cuprates.\ Contrarily to
the observations in Bi2212 and LaSrCuO, the maximum hole doping distribution 
$\Delta p$ found in YBCO is small enough to conclude that our YBCO samples
do not consist in interleaved regions with qualitatively different physical
properties (metal versus insulator, with or without pseudogap, etc). The
only remaining possibility for charge segregation or stripe-like scenarii to
apply inYBaCuO is then a dynamical process where the phase separation would
have a lifetime smaller than our timescale of observation.\ For the present
experiment, it corresponds to the inverse spectral width, typically 1 m$\sec 
$. Such dynamics might have been detected in inelastic neutron scattering
experiments, but the existing results are still under debate \cite{mook}\cite
{bourges}.

\bigskip Fig.\ Legend :

fig.1 : $^{89}$Y\ NMR shift $^{89}K$ for $H_{ext}\bot c$ for different
oxygen contents $y$ versus temperature.\ In the inset, $T_{c}$ for each
sample is plotted versus hole doping with the same symbol as the one used
for $K$.

fig.2\ : $^{89}$Y NMR Fourier Transform spectra are plotted at different
temperatures and dopings in YBa$_{2}$Cu$_{3}$O$_{6+y}$ for $H_{ext}\bot c$
(black solid lines) with arbitrary normalization. In each panel, a doping
scale gives the relation between the shift $^{89}K$ and hole doping $p$.
Dotted lines are simulated spectra (see text). At O$_{7}$:$T=120$ K, an
additional dashed line is plotted, which corresponds to the simulation made
using the dotted distribution of the upper panel of fig.3.

fig.3 : The full black dots represent $T_{c}$ versus oxygen content (upper
panel) and hole doping (lower panel) for YBa$_{2}$Cu$_{3}$O$_{6+y}$.
Distributions of oxygen content $P(y)$ and the corresponding distribution of
hole content ${\cal P}(p)$ were used to fit the spectra for $y=1$ and $y=0.6$
samples (see text).\ The dark and light gray enveloppes are obtained with a
minimal distribution of chemical shift and hyperfine coupling within
assumptions ({\it i)} and ({\it ii)} respectively. The dashed distribution
and the arrow shown in the lower panel represent the distributions measured
in Bi2212 \cite{Pan} and LaSrCuO \cite{Singer}.

*present adress : Dept of Physics, IIT Bombay 400076 India.


\begin{references}
\bibitem{emery}  V.J.\ Emery, S.A.\ Kivelson, H.Q.\ Lin, Phys.\ Rev.\ Lett. 
{\bf 64}, 475 (1990); J. Zaanen, O. Gunnarsson, Phys.\ Rev.\ B {\bf 40},
7391 (1989); V.J. Emery, and S.A. Kivelson, Physica C 209, 597 (1993).

\bibitem{NeutronsStripes}  J. M. Tranquada {\it et al.}, Nature {\bf 375},
561 (1995)

\bibitem{ChangPRB92}  J.-X. Liu {\it et al.}, Phys.\ Rev.\ Lett. {\bf 67},
2195 (1991); A.\ Chang {\it et al.}, Phys.\ Rev.\ B {\bf 46}, 5692 (1992)

\bibitem{Cren}  T.\ Cren {\it et al.}, Phys.\ Rev.\ Lett. {\bf 84}, 147
(2000)

\bibitem{Pan}  S.H.\ Pan {\it et al.}, Nature, {\bf 413}, 282 (2001)

\bibitem{howald}  C. Howald, P. Fournier, and A. Kapitulnik, Phys.\ Rev. B 
{\bf 64}, 100504 (2001)

\bibitem{LangSTM}  K.M.\ Lang {\it et al.}, Nature, {\bf 415}, 412 (2002)

\bibitem{edwards}  H.L.\ Edwards {\it et al.}, Phys.\ Rev.\ Lett. {\bf 75},
1387 (1995)

\bibitem{yeh}  N.-C.\ Yeh {\it et al.}, Phys.\ Rev.\ Lett. {\bf 87}, 087003
(2001)

\bibitem{Singer}  P.M. Singer, A.W. Hunt and T. Imai, Phys. Rev. Lett. {\bf %
88}, 047602 (2002)

\bibitem{FarellPRB}  D.E.\ Farrell {\it et al.}, Phys.\ Rev.\ B {\bf 36},
4025 (1987)

\bibitem{footnoteAlignement}  The orientation of the samples was chosen with 
$H_{ext}$ perpendicular to the cristallographic $c$ axis.\ In this
direction, the spectra are weakly affected by the small fraction of non
aligned polycristallite grains present in the sample batch.

\bibitem{Alloul89}  H.Alloul, T.Ohno and P.Mendels, Phys Rev.Lett. {\bf 63},
1700 (1989); H.\ Alloul {\it et al.}, Phys.\ Rev.\ Lett. {\bf 70}, 1171
(1993)

\bibitem{TallonPRB95}  J.L.\ Tallon {\it et al.}, Phys.\ Rev.\ B {\bf 51},
12911 (1995)

\bibitem{Alloul1990}  H.Alloul {\it et al.}, Physica Amsterdam C {\bf 171},
419 (1990)

\bibitem{footnote1}  The expected dipolar braodening from the nuclear Cu
moments is an order of magnitude smaller than the observed linewidth.

\bibitem{FootnoteOverdoping}  We however allow for a small local doping
distribution above $y=1$.\ To do so in the following analysis, we
extrapolate linearly the $K$ values obtained between $y=0.95$ and $y=1$ up
to $y=1.1$, as observed in Hg compounds \cite{bobroffPRL97}.

\bibitem{Haase}  J.\ Haase {\it et al.}, Physica Amsterdam C {\bf 341-348},
1727 (2000)

\bibitem{bobroffPRL97}  J.\ Bobroff {\it et al.}, Phys.\ Rev.\ Lett. {\bf 78}%
, 3757 (1997)

\bibitem{mook}  H.A. Mook and F. Dogan, Physica Amsterdam C {\bf 364-365},
553 (2001)

\bibitem{bourges}  P.\ Bourges {\it et al.}, Science {\bf 288}, 1234 (2000)
\end{references}
\end{document}